\documentclass[10pt,preprintnumbers,amsmath,amssymb,floatfix,twocolumn,prl]{revtex4} %revtex looks nice for cond-mat
\usepackage{graphicx}% Include figure files
\usepackage{dcolumn}% Align table columns on decimal point
\usepackage{bm}% bold math
\usepackage{longtable}
\usepackage{graphics}
\usepackage{amssymb}
\usepackage{amsmath}
\usepackage{xspace}
\usepackage{epsfig}
\newcommand {\etal}{{\it et al}. }
\newcommand {\etalc}{{\it et al}., }
\begin{document}
\title{Comparison of the %experimental
phase diagram of the half-filled layered
organic superconductors with the %theoretical
phase diagram of the
RVB theory of the Hubbard--Heisenberg model}
\author{B. J. Powell}
\email{powell@physics.uq.edu.au}
\author{Ross H. McKenzie}
\affiliation{Department of Physics, University of Queensland,
Brisbane, Queensland 4072, Australia}

%\pacs{87.15.-v, 82.35.Cd, 87.64.Je}

\begin{abstract}
We present an resonating valence bond (RVB) theory of
superconductivity for the Hubbard--Heisenberg model on an
anisotropic triangular lattice. We show that these calculations
are consistent with the observed phase diagram of the half-filled
layered organic superconductors, such as the $\beta$, $\beta$',
$\kappa$ and $\lambda$ phases of (BEDT-TTF)$_2X$
[bis(ethylenedithio)tetrathiafulvalene] and (BETS)$_2X$
[bis(ethylenedithio)tetraselenafulvalene]. We find a first order
transition from a Mott insulator to a $d_{x^2-y^2}$ superconductor
with a small superfluid stiffness and a pseudogap with
$d_{x^2-y^2}$ symmetry. The Mott--Hubbard transition can be driven
either by increasing the on-site Coulomb repulsion, $U$, or by
changing the anisotropy of the two hopping integrals, $t'/t$. Our
results suggest that the ratio $t'/t$ plays an important role in
determining the phase diagram of the organic superconductors.
\end{abstract}

\maketitle

Describing strongly correlated electronic systems is one of the
outstanding challenges of theoretical physics. In particular one
would like to understand if different model materials embody the
same underlying physics. The similarities between the cuprates and
the layered organics superconductors \cite{Ross_science,Kanoda}
suggest that similar physics may be realised in both classes of
materials. A powerful approach to chemically complex materials,
such as organic superconductors, is to define minimal models
\cite{Ross_review}, which can then be treated at various levels of
approximation \cite{Baskaran,FLEX}. In this Letter we take such an
approach. We argue that the observed phase diagram of the
half-filled layered organic superconductors ($\frac{1}{2}$LOS) is
well described by the RVB theory of the Hubbard--Heisenberg model.
Our theory reproduces the first order Mott transition and predicts
$d_{x^2-y^2}$ superconductivity \cite{disorder}, a small
superfluid stiffness \cite{constraints} and a pseudogap
\cite{pseudogap}.

%The many classes of organic superconductor include the
%quasi-one-dimensional Bechgaard salts and the
%quasi-two-dimensional layered organic superconductors
%\cite{Muller_review}.
Layered organic superconductors form several
crystal structures, some of which, such as the $\beta$, $\beta'$,
$\kappa$ and $\lambda$
phases are strongly dimerised, others, e.g., the %$\beta''$,
$\alpha$, $\beta''$ and $\theta$ phases are not. The chemical
composition of these materials is $D_2X$ where $D$ is an organic
donor molecule, for example BEDT-TTF (ET) or BETS, and $X$ is an
anion. Crystals consist of alternating layers of donor molecules
and anions \cite{Muller_review}. In both the dimerised and
undimerised salts the anion accepts one electron from a pair of
donor molecules which leads, at the level of band structure, to an
insulating anionic layer and a metallic donor layer. Quantum
chemistry suggests that the band structure of the undimerised
materials is well described by treating each donor molecule as a
site in a (quarter-filled) tight-binding model
\cite{Mernio_quater_filled}. %Thus the undimerised materials may be
%treated as quarter-filled with holes.
In the dimerised materials
the intra-dimer hopping integral is large enough that the band
structure can be described by a half-filled tight-binding model
with each site representing a dimer
\cite{Ross_review,constraints}.

$\frac{1}{2}$LOS display insulating, metallic, superconducting,
`bad metallic' and (possibly) pseudogap \cite{pseudogap} phases.
The nature of the superconducting state in $\frac{1}{2}$LOS is
controversial \cite{disorder}: the pairing is thought to be
singlet \cite{disorder}, but experiments have lead to both s-wave
and d-wave scenarios being proposed. Both phononic and
non-phononic pairing mechanisms have previously been considered
\cite{constraints,FLEX,Baskaran}. The superfluid stiffness
\cite{Pratt} is much smaller than is predicted by BCS theory but
is too large for fluctuations in the phase of the order parameter
to be important \cite{constraints}. %In this Letter we argue that many of the
%features of this phase diagram are well described by the RVB
%theory of superconductivity \cite{Anderson87,Laughlin,Zhang_gos}
%based on the Hubbard--Heisenberg Hamiltonian on an anisotropic
%triangular lattice.

\begin{figure}[!h]
    \centering
    \epsfig{figure=1a_phase_diagram_kX_data.eps, width=8cm,
    angle=0}
    \epsfig{figure=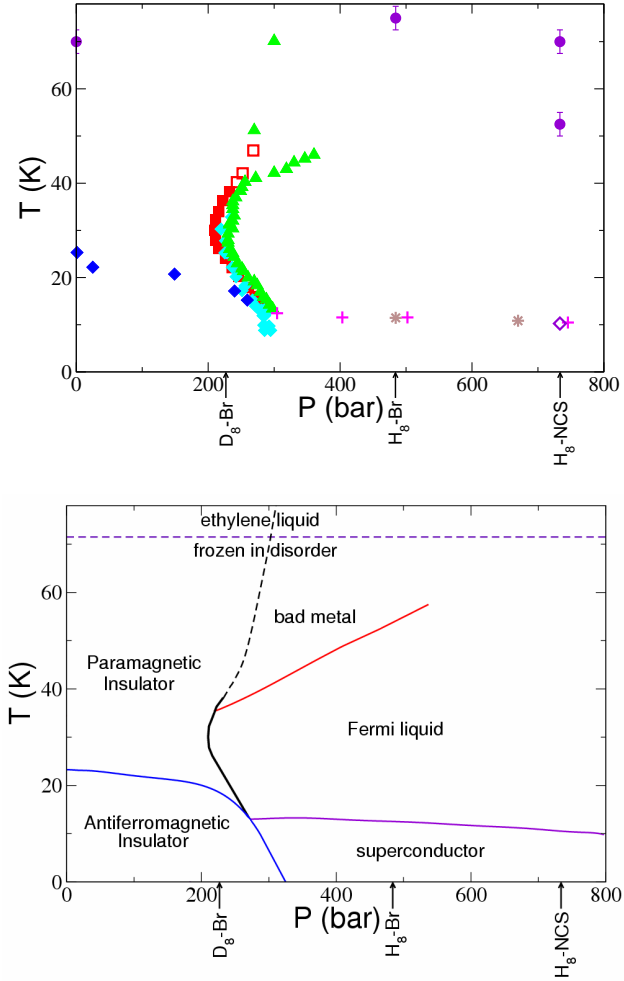, width=6.5cm, angle=270}
    \caption{(Color online.) The pressure-temperature phase diagram of $\kappa$-(ET)$_2X$.
    Top: data from $^1$H NMR and AC susceptibility (dark blue
    diamonds show the transition from a non-magnetic state to an antiferromagnetically
    ordered state, light blue diamonds show the metal-insulator transition \cite{Lefebvre}),
    magnetisation (pink pluses \cite{Schirber}), thermal expansion (filled
    purple circles \cite{Muller}),
    and resistivity (red squares \{filled indicates a first order Mott
    transition,
    empty indicates a crossover from insulating to metallic behaviours\}\cite{Kagawa},
    filled green triangles \cite{Limelette}, grey stars \cite{Taniguchi03} and
    open purple diamond \cite{Caulfield}).
    %This data comes from a number of different materials.
    We have
    offset the data to allow for the effect of `chemical
    pressure'. $P=0$ corresponds to ambient pressure for
    $X=$Cu[N(CN)$_2$]Cl \cite{Lefebvre,Schirber,Muller,Kagawa,Limelette}.
    The `chemical' pressure is indicated by the
    arrows on the abscissa. H$_8$-NCS $\Rightarrow X=$Cu(NCS)$_2$ \cite{Caulfield,Muller}
    and H$_8$-Br $\Rightarrow X=$Cu[N(CN)$_2$]Br  \cite{Taniguchi03,Muller}.
    D$_8$-Br
    indicates the effective chemical pressure of $X=$Cu[N(CN)$_2$]Br with the ET
    molecule fully deuterated \cite{Taniguchi03}.
    Bottom: a schematic version of the same diagram is shown.
    The glassy transition between the ethylene liquid and frozen in disorder
    phases results from conformational disorder in the organic molecule \cite{disorder,Muller}.
    %In the Hubbard and Hubbard--Heisenberg
    %models the antiferromagnetic and paramagnetic insulating phases are
    %Mott insulators, and the superconductivity and the bad metal phase arise
    %from strong electronic correlations.
    } \label{fig:phase_diagram}
\end{figure}

Fig. \ref{fig:phase_diagram} shows the phase diagram of
$\kappa$-(ET)$_2X$ as a function of pressure (both hydrostatic and
`chemical') and temperature. %We stress that
Other
$\frac{1}{2}$LOS have similar phase diagrams \cite{Muller_review}. %This diagram
%is extracted from a number of experiments
%\cite{Lefebvre,Schirber,Muller,Kagawa,Limelette,Taniguchi03,Caulfield}
%as described in the figure caption.
A simple explanation of this phase diagram is as follows
\cite{Kanoda,Ross_review}: there is a strong on-site Coulomb
repulsion, $U$, which causes the ambient pressure (Mott)
insulating state. The application of hydrostatic pressure or
varying the anion (often thought of as applying `chemical
pressure') reduces $U/W$, where $W$ is the bandwidth, and leads to
a superconducting state caused by strong electronic correlations.
The bad metal phase is due to somewhat localised electrons as one
crosses over from the Fermi liquid to the Mott insulator (which
does not require a phase transition
in these materials \cite{Kagawa,Limelette}). %In this Letter we will elaborate on this picture by presenting detailed
%calculations for a model which contains some of the features seen
%experimentally.

It has been argued that the Hubbard model on an anisotropic
triangular lattice is a minimal model for the layered organic
superconductors \cite{Ross_review}. A Dynamical Mean Field Theory
(DMFT) of the Hubbard model on a hypercubic lattice gives a good
quantitative description of the competition between the Mott
insulator, the bad metal and the Fermi liquid
\cite{Limelette,Merino00}. However, a mean field treatment of the
positive $U$ Hubbard model will not correctly describe the
materials as it neglects important spin correlations which arise
from superexchange. We therefore consider the Hubbard--Heisenberg
model, which can be derived \cite{Yu} from the Hubbard model in
the limit of large, but finite, $U$. The Hamiltonian is
%\begin{widetext}
\begin{eqnarray}
{\cal H} &=& \mu\sum_{i\sigma}\hat{n}_{i\sigma}
-t\sum_{\{ij\}\sigma}\hat{c}_{i\sigma}^\dagger\hat{c}_{j\sigma}
-t'\sum_{\langle
ij\rangle\sigma}\hat{c}_{i\sigma}^\dagger\hat{c}_{j\sigma} \notag \\
%\end{eqnarray}
%\begin{eqnarray}
&&+ J\sum_{\{ij\}}\hat{\bf S}_i\cdot\hat{\bf S}_j +J'\sum_{\langle
ij\rangle}\hat{\bf S}_i\cdot\hat{\bf S}_j +
U\sum_i\hat{n}_{i\uparrow}\hat{n}_{i\downarrow}\label{eqn:Ham}
\hspace{0.5cm}
\end{eqnarray}
%\end{widetext}
%\begin{eqnarray}
%{\cal H} &=& {\cal H}_t + {\cal H}_J + {\cal H}_U \label{eqn:Ham} %\\
%\end{eqnarray}
%where
%\begin{eqnarray}
%{\cal H}_t&=& \mu\sum_{i\sigma}\hat{n}_{i\sigma}
%-t\sum_{\{ij\}\sigma}\hat{c}_{i\sigma}^\dagger\hat{c}_{j\sigma}
%-t'\sum_{\langle
%ij\rangle\sigma}\hat{c}_{i\sigma}^\dagger\hat{c}_{j\sigma}
%\\
%\end{eqnarray}
%\begin{eqnarray}
%\cal H}_J&=& J\sum_{\{ij\}}\hat{\bf S}_i\cdot\hat{\bf S}_j
%+J'\sum_{\langle ij\rangle}\hat{\bf S}_i\cdot\hat{\bf S}_j
%\\
%\end{eqnarray}
%\begin{eqnarray}
%{\cal H}_U&=& U\sum_i\hat{n}_{i\uparrow}\hat{n}_{i\downarrow}
%\end{eqnarray}
where $\hat{c}_{i\sigma}^{(\dagger)}$ annihilates (creates) an
electron on site $i$ with spin $\sigma$, $\hat{\bf S}_i$ is the
Heisenberg spin operator, $\hat{n}_{i\sigma}$ is the number
operator, and $\{ij\}$ and $\langle ij\rangle$ indicate sums over
nearest neighbours and next nearest neighbours across one diagonal
only \cite{Ross_review} respectively. In principle $J=4t^2/U$ and
$J'=4t'^2/U$ to leading order due to superexchange. However, our
mean-field treatment will not correctly describe the
renormalisation of the bare parameters. Therefore we treat $t$,
$t'$, $J$, $J'$ and $U$ as independent parameters. To reduce our
parameter space we choose $J=t/3$ and $J'=t'^2/3t$ \cite{Jprime},
which correspond roughly to the values of $J$ and $J'$ extracted
from experiments on the insulating phase of the layered organics
\cite{Ross_review,Shimizu}. Thus in the calculations presented
below we only vary two parameters: $t'/t$ and $U/t$.

Our treatment of the Hubbard--Heisenberg Hamiltonian
(\ref{eqn:Ham}) is based on Anderson's RVB theory
\cite{Anderson87}. Although, the RVB wavefunction is a poor
approximation for the Heisenberg model on a square lattice, it was
recently shown that it is a good trial wavefunction for some
frustrated Heisenberg models \cite{Capriotti}. These models are
closely related to ours %. Based on electronic structure
%calculations and experiment
as many of the $\frac{1}{2}$LOS are expected
\cite{Ross_review,Shimizu} to have $J'/J=(t'/t)^2$ in the relevant
range. Further evidence that the RVB theory is a much better
theory for the triangular lattice than it is for the square
lattice comes from the critical value of the on site Coulomb
repulsion, $U_c$, at which the Mott transition occurs. The square
lattice is insulating for arbitrarily small values of $U$, whereas
we find that the RVB theory gives $U_c\simeq10.3t$. On the
isotropic triangular lattice exact diagonalisation of finite
lattices gives $U_c=12t$ \cite{Capone} and we find that
$U_c\simeq12.4t$ in the RVB theory.

Anderson's RVB state, $|RVB\rangle$, is given by performing a
Gutzwiller projection,
$\hat{P}_G=\sum_i(1-\alpha\hat{n}_{i\uparrow}\hat{n}_{i\downarrow})$,
on the BCS wavefunction, $|BCS\rangle$, i.e., $|RVB
\rangle=\hat{P}_G|BCS\rangle$. Here $\alpha$ is a variational
parameter which controls the fraction of doubly occupied sites,
$d$. A detailed analysis of the RVB theory of the
Hubbard--Heisenberg model on the square lattice was reported by
Gan \etal \cite{Zhang_gos}.

Following the spirit of Ref. \onlinecite{plain_vanilla} we make
the Gutzwiller approximation \cite{Fulde}, {\it viz}., $\langle
\hat{c}_{i\sigma}^\dagger\hat{c}_{j\sigma}\rangle_{RVB} = g_t
\langle \hat{c}_{i\sigma}^\dagger\hat{c}_{j\sigma}\rangle_{BCS}$
and $\langle \hat{\bf S}_i\cdot\hat{\bf S}_j \rangle_{RVB} = g_S
\langle \hat{\bf S}_i\cdot\hat{\bf S}_j\rangle_{BCS}$ %where
%$\langle{\cal O}\rangle_{RVB}\equiv\langle RVB|{\cal
%O}|RVB\rangle$ and $\langle{\cal O}\rangle_{BCS}\equiv\langle
%BCS|{\cal O}|BCS\rangle$.
where $\langle{\cal O}\rangle_{\psi}\equiv\langle\psi|{\cal
O}|\psi\rangle$. Counting arguments show that
\cite{Zhang_gos,Fulde} at half filling $g_t=8(1 - 2d)d$ and
$g_S=4(1-2d)^2$. The Gutzwiller approximation has several
advantages: its simplicity allows some analytic progress to be
made and allows one to consider infinite systems. %, it is
%equivalent to a slave-boson treatment and is exact in the limit of
%infinite dimensions \cite{MetznerVollhardt}.
However, the Gutzwiller approximation suppresses spin and charge
fluctuations in the Hubbard model \cite{Fulde}. We have already
sidestepped this problem somewhat by explicitly including the spin
exchange terms in the Hubbard--Heisenberg model. Our theory
produces a Mott insulating state that is a spin liquid rather than
the antiferromagnetic insulating state observed in most (but not
all \cite{Shimizu}) $\frac{1}{2}$LOS, however generalisation of
$|RVB\rangle$ to allow for antiferromagnetism should not
significantly alter the phase diagram. Clearly an important test
will be to project $|RVB\rangle$ onto the results of exact
diagonalisation of finite systems for the Hubbard model on the
anisotropic triangular lattice.

Making the Hartree--Fock--Gorkov approximation leads to two
coupled gap equations, $\Delta_{\bf k}=-\sum_{{\bf k}'}V_{{\bf
k}-{\bf k}'}\frac{\Delta_{{\bf k}'}}{2E_{{\bf k}'}}$ and
$\chi_{\bf k}=\tilde{\varepsilon}_{\bf k} - \sum_{{\bf k}'}V_{{\bf
k}-{\bf k}'}\frac{\chi_{{\bf k}'}}{2E_{{\bf k}'}}$
%\begin{eqnarray}
%\Delta_{\bf k}&=&-\sum_{{\bf k}'}V_{{\bf k}-{\bf k}'}\frac{\Delta_{{\bf k}'}}{2E_{{\bf k}'}} \label{eqn:Delta}
%%\\
%\end{eqnarray}
%\begin{eqnarray}
%\chi_{\bf k}&=&\tilde{\varepsilon}_{\bf k} - \sum_{{\bf
%k}'}V_{{\bf k}-{\bf k}'}\frac{\chi_{{\bf k}'}}{2E_{{\bf k}'}}
%\label{eqn:chi}
%\end{eqnarray}
where $V_{\bf k}=-\frac{3}{2}g_S[J(\cos k_x+\cos k_y) +
J'\cos(k_x+k_y)]$, $\tilde{\varepsilon}_{\bf
k}=\tilde{\mu}-g_t2[t(\cos k_x+\cos k_y)+ t'\cos(k_x+k_y)]$ and
$E_{\bf k} = \sqrt{[\tilde\varepsilon_{\bf k}+\chi_{\bf
k}]^2+|\Delta_{\bf k}|^2}$. $d$ is minimised variationally
%given by
%\begin{eqnarray}
%d= \frac{U+8\langle {\cal H}_t\rangle_{BCS} - 16\langle {\cal
%H}_J\rangle_{BCS}}{32\big(\langle {\cal H}_t\rangle_{BCS} -
%\langle {\cal H}_J\rangle_{BCS}\big)}
%\end{eqnarray}
and the renormalised chemical potential, $\tilde{\mu}$, is chosen
to ensure half-filling  \cite{Zhang_gos}.
%determined from the constraint that, at half filling,
%$1=\left\langle\sum_{i\sigma}\hat{n}_{i\sigma}\right\rangle_{BCS}/N=\sum_{\bf
%k}\left(1-\chi_{\bf k}/E_{\bf k}\right)$, where $N$ is the number
%of sites.
The two mean-fields are a Hartree--Fock term, $\chi_{\bf
k}=\sum_{{\bf k}'}V_{{\bf k}-{\bf k}'}\langle\hat{c}_{{\bf
k}'\uparrow}^\dagger\hat{c}_{{\bf k}'\uparrow}\rangle_{BCS}$, and
an anomalous term, $\Delta_{\bf k}=\sum_{{\bf k}'}V_{{\bf k}-{\bf
k}'}\langle\hat{c}_{{\bf k}'\uparrow}\hat{c}_{-{\bf
k}'\downarrow}\rangle_{BCS}$, where $\hat{c}_{{\bf k}\sigma}$ is
the Fourier transform of $\hat{c}_{{i}\sigma}$.

\begin{figure}
    \centering
    \epsfig{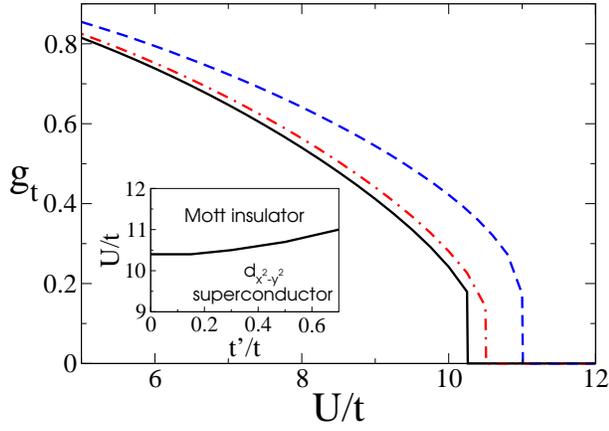}
    \caption{(Color online.) $g_t\sim Z$ as a
    function of the on-site Coulomb repulsion $U$ for various levels
    of values of $t'/t$, throughout $J=t/3$ and $J'=t'^2/3t$ \cite{Jprime}.
    Here we plot $t'=0.7t$ (dashed),
    $t'=0.3t$ (dot dashed) and $t'=0$ (solid). For each value of $t'/t$ we find a
    first order Mott-Hubbard transition at some critical value of
    the on-site Coulomb repulsion, $U_c$,
    from a superconducting state ($g_t\ne0$) to an insulating state ($g_t=0$).
    The inset shows the phase diagram for the model.
    Note that the fact that $U_c$ varies with $t'/t$ shows that
    increasing $t'/t$ can drive the Mott--Hubbard transition.
    %An artifact
    %of the Hubbard--Heisenberg model is that the gap remains finite
    %even as $U$ becomes small, however even within this approximation for
    %small $U$ the gap is so small (c.f Fig. \ref{fig:Delta_both})
    %that extremely small levels of disorder would completely suppress
    %superconductivity \cite{disorder}.
    } \label{fig:d_U}
\end{figure}

We solve the coupled gap equations %(\ref{eqn:Delta}) and (\ref{eqn:chi})
self consistently in reciprocal space on a
$120\times120$ mesh. We do not enforce any symmetry constraints on
the order parameters and we find that it has $d_{x^2-y^2}$
symmetry, this is the pairing symmetry most compatible with a
range of experiments on the layered organics \cite{disorder}.

\begin{figure}
    \centering
    \epsfig{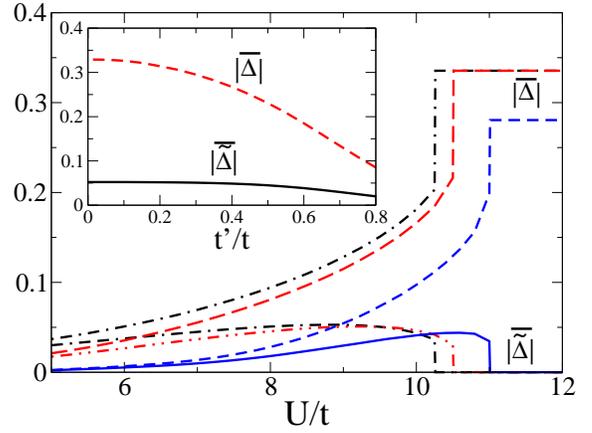}
    \caption{(Color online.) The mean gap, $\overline{|\Delta_{\bf k}|}$,
     and the mean superconducting
    order parameter, $\overline{|\widetilde{\Delta}_{\bf k}|}$,
    (both in units of $t$) as
    functions of the on-site Coulomb repulsion $U$ for various
    values
    of $t'/t$, throughout $J=t/3$ and $J'=t'^2/3t$ \cite{Jprime}.
    $\widetilde{\Delta}_{\bf k}=g_t\Delta_{\bf k}$
    and the bar indicates averaging over the
    Brillouin zone (i.e., $\overline{f_{\bf k}}=\sum_{\bf k}f_{\bf
    k}/\sum_{\bf k}$).
    Here we plot $t'=0.7t$ (blue: $\overline{|\widetilde{\Delta}_{\bf k}|}$ solid;
    $\overline{|\Delta_{\bf k}|}$ dashed),
    $t'=0.3t$ (red: $\overline{|\widetilde{\Delta}_{\bf k}|}$ double dot
    dashed;
    $\overline{|\Delta_{\bf k}|}$ long dashed) and
    $t'=0$ (black: $\overline{|\widetilde{\Delta}_{\bf k}|}$ dot
    double dashed;
    $\overline{|\Delta_{\bf k}|}$ dot dashed).
    The inset shows $\overline{|\Delta_{\bf k}|}$,
    (dashed) and $\overline{|\widetilde{\Delta}_{\bf k}|}$, (solid)
    against $t'/t$ with $J=t/3$, $J'=t'^2/3t$ and
    $U=9t$.}\label{fig:Delta_both}
\end{figure}

For the Hubbard model $g_t=Z$, the quasiparticle weight (the
factor by which many-body effects reduce the bandwidth and Drude
weight and enhance the effective mass, $m^*$). In Fig.
\ref{fig:d_U} we plot $g_t$ against $U$ for several values of
$t'/t$. For all values of $t'/t$ we find that at some critical
value, $U_c$, there is a first order transition from a
superconducting state ($g_t\ne0$) to an insulating state
($g_t=0$). This is consistent with the first order
superconductor-insulator transition observed experimentally in
$\frac{1}{2}$LOS \cite{Kagawa,Limelette}. Note that near the Mott
transition $g_t$ is reduced (and hence $m^*$ is enhanced by a
factor of 3 or 4), consistent with the large effective mass seen
in the layered organic superconductors close to the insulating
state \cite{constraints}. Previous weak coupling approaches
\cite{FLEX} do not capture this large mass renormalisation. $g_t$
is closely related to the reduction in the Drude weight due to
strong correlations. Fig. 2 is quantitatively similar to to exact
calculations of the Drude weight for the Hubbard model on an
isotropic triangular model (c.f., Fig. 4 of Ref.
\onlinecite{Capone}). The fact that $g_t<1$ leads to a reduced
superfluid stiffness \cite{Gan} as is observed in
$\frac{1}{2}$LOS. However, if we interpret our results in the
simplest manner \cite{plain_vanilla} they suggest that the most
correlated materials have the most strongly suppressed superfluid
stiffness which is the opposite trend to that found experimentally
\cite{constraints,Pratt}.

$\widetilde{\Delta}_{\bf k}=\sum_{{\bf k}'}V_{{\bf k}-{\bf
k}'}\langle\hat{c}_{{\bf k}'\uparrow}\hat{c}_{-{\bf
k}'\downarrow}\rangle_{RVB}=g_t\Delta_{\bf k}$ is the
superconducting order parameter. Fig. \ref{fig:Delta_both} shows
the mean of $|\Delta_{\bf k}|$ and $|\widetilde{\Delta}_{\bf k}|$
as functions of $U$ for several
values of $t'/t$. %Here the bar indicates averaging over the
%Brillouin zone and $\widetilde{\Delta}_{\bf k}=\sum_{{\bf
%k}'}V_{{\bf k}-{\bf k}'}\langle\hat{c}_{{\bf
%k}'\uparrow}\hat{c}_{-{\bf
%k}'\downarrow}\rangle_{RVB}=g_t\Delta_{\bf k}$.
$\widetilde{\Delta}_{\bf k}\ne\Delta_{\bf k}$ indicates %that there is
a pseudogap, which is predicted to be largest near the Mott
transition %(as is seen experimentally)
and have $d_{x^2-y^2}$ symmetry. The angle dependence of the
pseudogap could be measured by angle resolved photoemission or
angle resolved magnetoresistance oscillations. We plot the mean of
$|\Delta_{\bf k}|$ and $|\widetilde{\Delta}_{\bf k}|$ as functions
of $t'/t$ for fixed \ref{fig:Delta_both}. Varying $t'/t$ can lead
to a suppression of superconductivity and can even drive the Mott
transition as can be seen from the phase diagram of the model
(inset to Fig. \ref{fig:d_U}).

Our results suggest that the effects of the anisotropy of the
triangular lattice are important for the organic superconductors.
Most importantly we suggest that the large value of $t'/t$ and
hence of $J'/J$ stabilises the RVB state in $\frac{1}{2}$LOS
\cite{Capriotti}. The RVB state naturally explains the first order
transition between the Mott insulator and a $d_{x^2-y^2}$
superconductor. Further, our results suggest that the simple
picture \cite{Ross_review,Kanoda} in which the only role of
hydrostatic and `chemical' pressure is to vary $U/W$ is not
sufficient to explain the phase diagram of $\frac{1}{2}$LOS. It
appears that the value of $t'/t$, and hence $J'/J$ also plays a
crucial role in determining the behaviour of these materials.
$t'/t$ controls the degree of nesting of the Fermi surface and
therefore directly controls the stability of the Mott insulator,
whereas $J'/J$ determines the pairwise potential $V_{\bf k}$ which
controls the stability of the superconducting phase. This is why
variations in $t'/t$ can even drive the Mott transition at a fixed
$U$. Clearly, the physics of the anisotropic triangular lattice is
qualitatively different from that of square lattice.

Combining our results with those from DMFT studies of the Hubbard
model allows one to reproduce the main features of the phase
diagram of $\frac{1}{2}$LOS (Fig. \ref{fig:phase_diagram}). This
is consistent with the claim \cite{Ross_review} that the Hubbard
model is the minimal model for $\frac{1}{2}$LOS. (We stress that
this theory is not applicable to the quarter-filled layered
organic superconductors \cite{Mernio_quater_filled} or the
Bechgaard salts \cite{Muller_review}.) However, caution is
required here. Although calculations (e.g., \cite{Limelette})
based on the Hubbard model can give good quantitative agreement
with experiments on $\frac{1}{2}$LOS one does not know {\it a
priori} how to map the experimental parameter space (pressure,
temperature and chemical composition) onto the theoretical
parameter space ($t'/t$, $U/t$, etc.). %To discover whether the
%Hubbard model correctly describes $\frac{1}{2}$LOS one must map
%the experimental parameter space onto the theoretical parameter
%space and show that such a mapping connects the equivalent
%experimental and theoretical phases.
Therefore an outstanding problem is to discover whether quantum
chemistry predicts the large variations of the Hubbard parameters
with pressure required for quantitative agreement with experiment.

We have presented an RVB theory of the Hubbard--Heisenberg model
on the anisotropic triangular lattice. We argued that the RVB
state may be a good trial wavefunction for $\frac{1}{2}$LOS
because the values of $J'/J$ deduced from quantum chemistry and
experiment are comparable to those for which the RVB state appears
to be a good approximation. Our calculations show a first order
Mott--Hubbard transition from an insulating state to a
d$_{x^2-y^2}$ superconductor. A similar first order Mott
transition is seen in experiments on $\frac{1}{2}$LOS. The
Mott--Hubbard transition can be driven by increasing {\it either}
$U/t$ or $t'/t$. Further, at a fixed $U$, superconductivity is
strongly suppressed by increasing $t'/t$. This suggests that the
value of $t'/t$ may be more important in the layered organic
superconductors than has previously been appreciated. The
superconducting state has a reduced superfluid stiffness as is
observed in the $\frac{1}{2}$LOS. The RVB theory predicts that
there is a pseudogap with $d_{x^2-y^2}$ symmetry.

Note added: after completing this work we became aware of some
similar results obtained by Gan \etal \cite{Gan} and Liu \etal
\cite{Liu}.

It is a pleasure to thank J.P. Barjaktarevic, N. Bonesteel, J.
Fjaerestad, J.B. Marston, A.P. Micolich, F.L. Pratt, P. Wzietek,
and K. Yang for stimulating conversations and Brown University
(BJP), Oxford University and Rutherford Appleton Laboratory for
hospitality. Numerics were enabled by an APAC grant. % from the
%Australian Partnership for Advanced Computing.
This work was funded by the Australian Research Council.


\begin{thebibliography}{99}

%\bibitem{deSoto}
%de Soto.

\bibitem{Kanoda}
K. Kanoda, Physica C {\bf 282-287}, 299 (1997).

\bibitem{Ross_science}
R.H. McKenzie, Science {\bf 278}, 820 (1997).

\bibitem{Ross_review}
%H. Kino and H. Fukuyama, J. Phys. Soc. Jpn. {\bf65}, 2158 (1996);
R.H. McKenzie, Comments Condens. Matter Phys. {\bf 18}, 309
(1998).

\bibitem{FLEX} J. Schmalian, Phys. Rev. Lett. {\bf81} 4232 (1998);
H. Kino and H. Kontani, J. Phys. Soc. Japan {\bf67} 3691 (1998);
H. Kondo and T. Moriya, J. Phys. Soc. Japan {\bf67} 3695 (1998);
T. Jujo, S. Koikegami and K. Yamada,  J. Phys. Soc. Japan {\bf68}
1331 (1999); R. Louati \etalc Synth. Met. {\bf103} 1857 (1999); M.
Vojta and E. Dagotto, Phys. Rev. B {\bf59} 713 (1999); H. Kontani,
Phys. Rev. B {\bf67}, 180503 (2003).

\bibitem{Baskaran} G. Baskaran, Phys. Rev. Lett. {\bf90}, 197007 (2003).

\bibitem{disorder}
B.J. Powell and R.H. McKenzie, Phys. Rev. B {\bf 69}, 024519
(2004).

\bibitem{constraints}
B.J. Powell and R.H. McKenzie, J. Phys.: Condens. Matter. {\bf
16}, L367 (2004) and references therein.

\bibitem{pseudogap}
K. Miyagawa, A. Kawamoto, and K. Kanoda, Phys. Rev. Lett. {\bf89},
017003 (2002).

\bibitem{Muller_review}
%For a recent review see %M. Lang and J. M\"uller, ``{\it Organic
%superconductors}" in ``{\it The Physics of Superconductors -
%Vol.2}", K.-H. Bennemann, J.B. Ketterson (Eds.), Springer-Verlag
%(2003) or
T. Ishiguro, K. Yamaji, and G. Saito, {\it Organic
Superconductors} (Springer Verlag, Heidelberg, 1998).

\bibitem{Mernio_quater_filled}
J. Merino and R.H. McKenzie, Phys. Rev. Lett. {\bf87}, 237002
(2001).

\bibitem{Pratt}
F. L. Pratt \etalc %, S.J. Blundell, I.M. Marshall, T. Lancaster,
%S.L. Lee, A. Drew, U. Divakar, H. Matsui and N. Toyota,
Polyhedron {\bf 22} 2307 (2003).

\bibitem{Lefebvre}
S. Lefebvre \etalc %, P. Wzietek, S. Brown, C. Bourbonnias, D. J\'erome, C.
%M\'ezi\`ere, M. Fourmigu\'e, and P. Batail,
Phys. Rev. Lett. {\bf
85}, 5420 (2000).

\bibitem{Schirber}
J.E Schirber \etalc %, D.L. Overmyer, K.D. Carlson, J.M. Williams, A.M.
%Kini, H. Hau Wang, H.A. Charlier, B.J. Love, D.M. Watkins and G.
%A. Yaconi,
Phys. Rev. B {\bf44}, 4666 (1991).

\bibitem{Muller}
J. M\"uller \etalc %, M. Lang, F. Steglich, J.A. Schlueter, A.M. Kini, and
%T. Sasaki,
Phys. Rev. B {\bf 65}, 144521 (2002).

%\bibitem{Taniguchi99}
%Taniguchi A. Kawamoto, and K. Kanoda, Phys. Rev. B {\bf 59}, 8424
%(1999)

\bibitem{Kagawa}
F. Kagawa, T. Itou, K. Miyagawa and K. Kanoda, Phys. Rev. B
{\bf69}, 064511 (2004).

\bibitem{Limelette}
P. Limelette \etalc %, P. Wzietek, S. Florens, A. Georges, T.A. Costi, C.
%Pasquier, D. J\'erome, C. M\'ezi\`ere and P. Batail,
Phys. Rev. Lett. {\bf91}, 016401 (2003).

\bibitem{Taniguchi03}
H. Taniguchi, K. Kanoda, and A. Kawamoto, Phys. Rev. B {\bf 67},
014510 (2003).

\bibitem{Caulfield}
J. Caulfield \etalc %, W. Lubczynski, F.L. Pratt, J. Singleton, D.Y.K. Ko,
%W. Hayes, M. Kurmoo and P. Day,
J. Phys.: Condens. Matter. {\bf 6}, 2911 (1994).

\bibitem{Merino00}
J. Merino and R.H. McKenzie, Phys. Rev. B {\bf61}, 7996 (2000).

\bibitem{Yu}
Y. Yu, cond-mat/0303501.

\bibitem{Jprime} This choice of $J$ and $J'$ ensures that
$J'/J=(t'/t)^2$ as required physically to leading order in $t^2/U$
and $t'^2/U$.

\bibitem{Shimizu}
Y. Shimizu \etalc %, K. Miyagawa, K. Kanoda, M. Maesato and G. Saito,
Phys. Rev. Lett. {\bf 91}, 107001 (2003); W. Zheng, R.R.P. Singh,
R.H. McKenzie and R. Coldea, cond-mat/0410381.

\bibitem{Anderson87}
P.W.~Anderson, Science {\bf 235}, 1196 (1987); F.C. Zhang, C.
Gross, T.M. Rice and H. Shiba, Supercond. Sci. Technol. {\bf 1},
36 (1988).

\bibitem{Capriotti}
L. Cappriotti, F. Becca, A. Parola and S. Sorella, Phys. Rev.
Lett. {\bf 87} 097201 (2001); S. Yunoki and S. Sorella, {\it
ibid}. {\bf92}, 157003 (2004).

\bibitem{Capone} M. Capone, L. Capriotti, F. Becca and S. Caprara,
Phys. Rev. B {\bf63}, 085104 (2001).

\bibitem{Zhang_gos}
J.Y. Gan, F.C. Zhang and Z.B. Su, cond-mat/0308398.



%\bibitem{Laughlin}
%R. Laughlin, cond-mat/0209269; F. C. Zhang, Phy. Rev. Lett., {\bf
%90}, 207002 (2003).

\bibitem{plain_vanilla} P.W. Anderson \etalc J. Phys.: Condens. Matter {\bf 16} R755 (2004).


\bibitem{Fulde}
For a review see P. Fulde, {\it Electron Correlations in Molecules
and Solids} (Springer Verlag, Heidelberg, 1991).

%\bibitem{MetznerVollhardt}
%W. Metzner and D. Vollhardt, Phys. Rev. Lett. {\bf 62}, 324
%(1989).

\bibitem{Gan}
J.Y. Gan, Y. Chen, Z.B. Su, and F.C. Zhang, cond-mat/0409482.

\bibitem{Liu}
J. Liu, J. Schmalian, and N. Trivedi, cond-mat/0411044.

\end{thebibliography}
\end{document}